\documentclass[prl,twocolumn,aps,amssymb,footinbib,showpacs]{revtex4-1}
\usepackage{amssymb}
\usepackage{amsmath}
\usepackage{times}
\usepackage{latexsym}
\usepackage{graphicx}
\usepackage{color}
\usepackage{bm}
\usepackage{subfigure}
\begin{document}
\title{Majorana Fermions: Anholonomy of Bound States}
\author{\bf Sourin Das$^{1,2}$ and  Indubala I Satija$^3$ }
\email{sdas@physics.du.ac.in, isatija@gmu.edu}
\affiliation{$^1$ Department of Physics and Astrophysics, University of Delhi, Delhi 110007, India
\\
$^2$ Max Planck Institute for the Physics of Complex Systems, N{\" o}thnitzer Strasse 38, 01187 Dresden, Germany 
\\
$^3$ Department of Physics and Astronomy, George Mason University , Fairfax, VA 22030, USA}
\date{\today}
\begin{abstract}
{Majorana bound states appearing in 1-D $p$-wave superconductor ($\cal{PWS}$) are found to result in exotic quantum holonomy of both eigenvalues and the  eigenstates. Induced by a degeneracy hidden in complex Bloch vector space, Majorana states are identified with  a pair of exceptional point ($\cal{EP}$) singularities. Characterized by a collapse of the vector space, these singularities are defects in Hilbert space that lead to M$\ddot{\rm o}$bius   strip-like structure of the eigenspace and singular quantum metric. The topological phase transition in the language of $\cal{EP}$ is marked by one of the two exception point singularity degenerating to a degeneracy point with non singular quantum metric. This may provide an elegant and useful framework to characterize the topological aspect of Majorana fermions and the topological phase transition.}
\end{abstract}

\pacs{03.75.Ss,03.75.Mn,42.50.Lc,73.43.Nq}
\maketitle
{\it{{\underline{Introduction}}:-}} Presently, Majorana fermions occupy a central place in many vibrant frontiers of modern physics. These include neutrino physics, supersymmetry, dark 
matter, and superconductivity which offer possibilities where Majorana Fermions may be found\cite {Wil}. Following a pioneering work by Kitaev \cite{Kitaev},
the one-dimensional $p$-wave superconducting quantum wire has emerged as an ideal system for theoretical \cite {Mftheo2,Mftheo3,Mftheo4,Mftheo5} and experimental  \cite{Mfexpt0,Mfexpt1,Mfexpt2} exploration of Majorana Fermions. Existence of a zero energy Majorana bound state protected by a gap appearing for an open chain  boundary  ( $\cal{OBC})$ and  which disappears for periodic boundary condition ($\cal{PBC}$) was shown to be a hallmark of this topologically  nontrivial phase which was characterized by a  $Z_2$-index \cite{Kitaev}. 

In this Letter, we lay down an independent  route for identifying the presence of Majorana bound states in terms of exceptional points ($\cal{EP}$) \cite{heiss, Keck,Korsch, BerryNH}. It is shown that degenerate zero energy sub-gap states, which can be identified with the Majorana bound states,  appear for the case of $\cal{PBC}$ provided one extends the Hilbert space to include complex valued Bloch wave vectors (call it $k=k_r+i \, k_i$  where $k_r,\, k_i \in \mathbb{R}$) leading to non-Hermitian Hamiltonian description of the system. The zero energy eigenstates of this  non-Hermitian Hamiltonian embedded in the $k_r-k_i$ plane are shown to have square root singularity, encircling which once results in swapping of eigenvectors and encircling twice results in the eigenvectors acquiring a nontrivial phase factor.\footnote{It should noted that based on our idea of describing  Majorana zero mode as $\cal{EP}$ its application to higher dimensions has already been published in EPL,110, 67005 (2015) by I Mandal}

In contrast to the Berry phase which is an anholonomy in the eigenvector space, the non-Hermitian anholonomy can exist both in the eigenvalue and in the eigenvector space, exhibiting branch point singularity in the complex $k$-plane. Such phenomena are popularly referred  to as the exotic quantum holonomy \cite{exotic}. 

 Seminal work by Bender \cite{Bender2,Bender1} had set a stage for the importance of non-Hermitian systems as it was argued that the complex domain being huge compared to the real domain  and thus opens new possibilities such as new types of unitary evolutions. In a different analysis, analytic continuation of the energy in complex plane provided new insight about topological edge modes in quantum Hall systems \cite{Hatsugai}.  Following a similar line of thought we establish a direct connection between the $\cal EP$ and the Majorana fermions. 

 The letter is organized in the following fashion:  {\it{(a)}} we first show that there are two pairs of zero energy solutions when the Hamiltonian is written in complex $k$-basis. Of these we identify the pair which faithfully represents the Majorana zero modes,  {\it{(b)}} we then expand the Hamiltonian around the these zero energy solution to establish them as $\cal{EP}$,    {\it{(c)}} next,  we  characterize the zero energy solutions in terms of its Berry phase and quantum geometry which reveal complementary information regarding the $\cal EP$s, {\it{(d)}} finally we show that the quantum phase transition in the model is characterized by disappearance of one of the $\cal{EP}$ from the pair exactly at the transition point. Finally we also propose a topological quantum number which is expressed in terms of $\cal{EP}$s and provides the count of Majoranas in the topological phase.\\
 {\it{{\underline{Hamiltonian and its Jordan form at zero energy solutions}}:-}}We begin with Kitaev's Hamiltonian  for the 1-D $\cal{PWS}$ given by
\begin{equation}
H=\sum  \frac{w}{2} \, c_{n+1}^{\dagger}c_n^{} + \Delta\, c_{n+1}^{\dagger} c^{\dagger}_n + \textrm{c.c.} + \mu\, (c_n^{\dagger} c_n^{} -1/2) ~.
\end{equation}
Here $w$ is the nearest-neighbor hopping amplitude, $\Delta=\vert \Delta \vert \exp{(i \theta)}$ the superconducting gap function, and $\mu$ the on-site chemical potential. From here on we we will express $\vert \Delta \vert $ and $ \mu$ in units of $w$, i.e, set $w=1$. A quantum phase transition occurs as one tunes the chemical potential $\mu$ from $\mu > 1$ (topologically trivial) to $\mu < 1$ (topological nontrivial) with $\mu=1$ being the point of transition \cite{Kitaev}.
In the Bogoliubov basis, the system reduces to a collection of  $2$-level systems for the case $\cal{PBC}$ given by,
\begin{equation}
H(k) = (\cos k - \mu)\, \sigma_y -\vert \Delta \vert  \sin k \, \sigma_x~,
\label{twolevel}
\end{equation}
Where $\sigma_x $ and $\sigma_y$ are the Pauli matrices. The energy spectrum, labeled by the Bloch index $k$ is given by,
\begin{equation} 
E_k^{\pm} = \pm \sqrt{ ( \mu-\cos k)^2 + \vert \Delta\vert^2  \sin ^ 2 k}  ~. 
\end{equation}
The energy spectrum of the system is in general gaped that vanishes at a critical value,  $\mu_c = 1$. Kitaev has shown that for  $\mu < 1$ the system has a two fold degenerate manybody ground state or no degeneracy depending upon if we impose $\cal{OBC}$ or $\cal{PBC}$ hence establishing it as a topological  phase. He also showed that the degeneracy is directly related to appearance of Majorana mode at each of the two ends of the 1-D $\cal{PWS}$ with $\cal{OBC}$. In this letter we show an independent route to this topological degeneracy within the $\cal{PBC}$ by extending the $k$-space to include complex eigenvalues though this renders the $H(k)$ non-Hermitian. 

In this approach the degeneracy appears purely in the complex $k$-plane. To analyze these degeneracies further lets us rewrite Eq.(\ref{twolevel}) in  an explicit non-Hermitian  matrix form 
\begin{eqnarray}
{H} (k) &=& \left( \begin{array}{cc} 0 & A_k\\ B_k& 0\\ \end{array}\right)~, 
\label{NH}
\end{eqnarray}
where $A_k= -a_k + i b_k ,B_k = -a_k -i b_k $ and $a_k=  \vert \Delta \vert \sin k$, $b_k=\mu - \cos k$ are complex quantities when $k$ is complex.
From here on we will assume $\Delta$ to be real without loss of generality 
as the phase $\theta$ does not have any physical consequence. The eigenvalues of $H(k)$ are given by  $E_{k}^{\pm}=\pm \sqrt{A_k}  
\sqrt{B_k}$ and the zero energy solution corresponds to either $A_k=0$ or  $B_k=0$. One can immediately see from  Eq.(\ref{NH})  that  
the $H(k)$ for the zero energy sector either takes the $2\times2$ Jordan normal form \cite{jordan} for $B_k=0, A_k \neq 0$ or a matrix which can be reduced to the   
$2\times2$ Jordan normal form by a similarity transformation for $A_k=0, B_k \neq 0$. The solutions corresponding to $A_k=0$ and  $B_k=0$ in the complex $k \equiv k_0=k_r+ik_i$ are given by 
\begin{eqnarray}
e^{i \,  k_{0}^{\pm,A}}& = &\frac{\mu \pm\sqrt{ \mu^2-(1-  { \vert \Delta \vert^2)}}}{1-  {\vert \Delta \vert }}, \\
e^{i \,  k_{0}^{\pm,B}}& = &\frac{\mu \pm\sqrt{ \mu^2-(1-  { \vert \Delta \vert^2)}}}{1+  {\vert \Delta \vert }}~.
\label{ising0}
\end{eqnarray}

We note that these two solutions are connected by the relation $e^{i \,  k_{0}^{\pm,A}}\, e^{i \,  k_{0}^{\pm,B}}$=$1$. The crucial physical insight which helps us to choose one of these two pairs of solutions as a true representative for the Majorana state appearing in the original problem discussed by Kitaev \cite{Kitaev} lies in the fact that the Majorana  solution found by Kitaev localizes at the boundary from being delocalized as one enters the topological phase. In the context of our approach involving complex $k$, this amounts to having imaginary part of momenta 
to be positive definite in the topological phase. It is straight forward to check that only $B_k=0$ solution satisfy this condition for the physically relevant part of the parameter space given by positive values of the chemical potential $ \mu \geq 0$ in the topological window $1 > \mu \geq 0$. We have presented a summary of this fact in Table~\ref{table1}. Since our aim is to estimable that Majorana zero modes in Kitaev's model can be identified with $\cal{EP}$s in the complex $k$ plane, hence we will focus only on $B_k=0$ solutions for the rest of the paper.
\footnote{For the unphysical case of negative $\mu$, the topological window for $\mu$ is given by $-1< \mu \leq 0$ and in this window the correct identification of Majorana solution corresponds to the  $A_k=0$ case.}.

 \begin{table}
\begin{tabular}{ c | c | c| c }
 $k_{0}^{A_+}$\,\,  & $ k_{0}^{A_-}$\,\, & $ k_{0}^{B_+}$ \,\, & $k_{0}^{B_-}$  
 \\ \hline
 $-i \ln\frac{1+{\vert \Delta \vert }}{1-{\vert \Delta \vert}} $
 \,\,  & 
 $-i  \frac{\epsilon}{{\vert \Delta \vert}}$ \,\, 
 & $ i \frac{\epsilon}{{\vert \Delta \vert}}$ \,\, 
 &  $i\ln\frac{1+{\vert \Delta \vert }}{1-{\vert \Delta \vert}}$  \,\,
 \label{Table1}
\end{tabular}
\caption{Here we have tabulated a small $\epsilon$ expansion ($\mu = 1-\epsilon$) for two pairs of zero energy momenta corresponding to $A_k=0, B_k \neq 0$ and $A_k\neq0, B_k=0$ up to leading order in $\epsilon$. The table clearly show that $B_k=0 (A_k=0)$ momenta undergoes a localization (delocalization) transition as $\mu$ cross unity starting from the $\mu >1$ side.} \label{table1}
\end{table}

Note that the Jordan form corresponds to matrices which are tridiagonal and they can not be diagonalized  by similarity transformations. The $2\times2$ Jordan block has two degenerate eigenvalues but its eigenstates collapse to a single state and hence do not span a  two-dimensional space. Therefore, in the manifold of eigenstates parametrized by complex valued $k$, $k=k_0^\pm$ represents a topological defect as the collapsed eigenspace corresponding to $k_0^\pm$ can not be continuously
 deformed to the two-dimensional eigenspace corresponding to rest of the values of $k$. These topological defects manifest themselves as square-root singularities as we expand eigenvector and eigenvalue of $H(k)$ about $k_0^\pm$ leading to a M$\ddot{\rm o}$bius strip like topology. Such singularities are popularly referred to as exceptional point ($\cal{EP}$) \cite{heiss,rotter} or non-Hermitian degeneracies \cite{BerryNH,BerryNH1,BerryNH2}. \\
{\it{{\underline{Eigenbasis for Kitaev's Hamiltonian at the $\cal{EP}$s}}:-}} Eigenstates spanning the Hilbert space of non-Hermitian $H(k)$ as in Eq.(\ref{NH}) are the bi-orthogonal vectors given by
\begin{equation}
 | ~R,\pm~\rangle=1/\sqrt{2} \left(\pm \sqrt{A_k/B_k}, 1\right)~,
 \end{equation} 
where the respective dual vectors are defined as  
\begin{equation}
\langle ~L,\pm~|=1/\sqrt{2} \left(\pm \sqrt{B_k/A_k}, 1\right)~,
\end{equation}
corresponding to  $E_{k}^{\pm}$. Here $L,R$ refer to left and right eigenvectors. These vectors  satisfy the bi-orthonormalization relation \cite{garrison,brody}
\begin{eqnarray}
\langle ~L,\pm~|~R,\pm~\rangle &=& 1  \quad; \quad
\langle ~L,\pm~|~R,\mp~\rangle=0~. 
\label{NC}
\end{eqnarray}
Though such a normalization procedure works very well for non-Hermitian Hamiltonian in general but it runs into trouble at the exceptional point due to the nontrivial topology associated with the $\cal{EP}$. We note that, up to overall  normalization and phases, for $B_k=0$, $|~R,\pm~\rangle=(1,0)$ and $\langle ~L,\pm~|=(0,1)$.  Hence tuning to the degeneracy point results  in the collapse of two-dimensional Hilbert space to a one-dimensional space as $| ~R,+~\rangle$  and $| ~R,-~\rangle$  become parallel to each other at this point. This is nothing but a manifestation of the non-diagonalizability of $H(k)$ at the exceptional point. Also note that now the dual vectors at $\cal{EP}$ has become mutually orthogonal, i.e. $\langle ~L,\pm~| ~R,\pm~\rangle = 0$ and they no more comply to the normalization defined in Eq.(\ref{NC}). 

To study the behavior of the wavefunction in the vicinity of the exceptional point, we need to perform a  systematic  perturbative expansion which allows for an expansion of $H(k)$ around the $\cal{EP}$. The first hurdle in this direction is to identify the eigenvectors of the unperturbed Hamiltonian, i.e. the Hamiltonian at the $\cal{EP}$ which serve as a  natural basis for performing the perturbation theory. Due to the collapse to the Hilbert at $\cal{EP}$ we have to first engineer a consistent way of reconstructing a two-dimensional Hilbert space around the $\cal{EP}$.  The $B_k=0$ degeneracy condition leads to a pair of $\cal{EP}$ at $k=k_0^\pm$ where we will further focus only on $k=k_0^+$ (call it $k_0$ henceforth) as the perturbation theory around both $\cal{EP}$s have the same analytic form. For simplifying notation we rename $\langle ~L,\pm~|$ and $| ~R,\pm~\rangle$ at $k=k_0^+$ as 
$\langle ~L,\cal{EP}~|$ and $| ~R,\cal{EP}~\rangle$. Note that the $\pm$ sign is redundant owing to the collapse of Hilbert space at the $\cal{EP}$. Following Ref.~\cite{brody} an associated Jordan vector (call it  $| ~R(a),\cal{EP}~\rangle$) is identified which essentially facilitates a systematic  perturbative expansion around $\cal{EP}$. Now a two-dimensional  space can be identified which is spanned by the following two linearly independent vectors,
\begin{equation}
 | ~R,{\cal{EP}}~\rangle=(1,0) ~,~  
 | ~R(a),{\cal{EP}}~\rangle=(-1/2 a_{k_0} ) (0,1)~.
 \end{equation}
  The corresponding  duals can be identified as,
  \begin{equation}
  \langle ~L,{\cal{EP}}~|=(-2 a_{k_0} ) (0,1) ~,~
\langle ~L(a),{\cal{EP}}~|=(1,0)~,
\end{equation}
which are subjected to the new set of orthonormalization condition
 \begin{eqnarray*}
  \langle ~L,{\cal{EP}}~|~R(a),{\cal{EP}}~\rangle &=&1 ~,~
\langle ~L(a),{\cal{EP}}~|~R,{\cal{EP}}~\rangle=1\\
\langle ~L,{\cal{EP}}~|~R,{\cal{EP}}~\rangle&=&0 ~, ~ \langle ~L(a),{\cal{EP}}~|~R(a),{\cal{EP}}~\rangle=0. 
\end{eqnarray*}
Note that these orthogonality and normalization conditions are distinct from that defined in Eq.(\ref{NC}) which are valid for a general diagonalizable non-Hermitian $2\times2$ Hamiltonian. \\
{\it{{\underline{Expansion of Kitaev's Hamiltonian around $\cal{EP}$s}}:-}}  A systematic expansion of $H(k)$ around the $\cal{EP}$ to leading order in $k-k_0$ which confirms to a form $H(k)=H(k_0)+H^\prime~(k-k_0)$ is given by
\begin{eqnarray}
{H}(k)&=& -2 a_{k_0} \left( \begin{array}{cc} 0 & 0\\ 1& 0\\ \end{array}\right) +\left( \begin{array}{cc} 0 & A_k^\prime\\ B_b^\prime& 0\\ \end{array}\right) (k-k_0)~, 
\label{CP}
\end{eqnarray}
where  $A_k^\prime=-\mu \Delta + i (1+\Delta^2) \sin k_0$, $B_k^\prime=-\mu \Delta - i (1-\Delta^2) \sin k_0$ are first derivatives of $A_k$, $B_k$ at $k=k_0$. Note that the leading correction $H^\prime$ to the non-diagonalizable Hamiltonian at ${\cal{EP}}$  is a diagonalizable matrix. The leading order expansion of eigenstates and eigenvalues in $k-k_0$ consistent with the Schr$\ddot{\rm o}$dinger equation for the form of $H(k)$ as in  Eq.(\ref{CP}) is given by 
\begin{eqnarray}
|~\pm,\theta~\rangle&=& \frac{1}{N}\{|~R,{\cal{EP}}~\rangle + \zeta_{\pm}\, \sqrt {r}\, |~R(a),{\cal{EP}}~\rangle ~e^{i \frac {\theta}{2}}\}~,
 \\
E_\pm(\theta)&=&\pm~ \zeta\, \sqrt {r}~e^{i \frac{\theta}{2}}~, 
\label{eigenstate}
\end{eqnarray}
where $k-k_0=r e^{i \theta}$, $\zeta_{\pm}=\pm \sqrt{ \langle~L,{\cal{EP}}~|~H^\prime |~R,~{\cal{EP}}~\rangle}=\sqrt{-2 \Delta \sin k_0 (-\mu \Delta - i (1-\Delta^2) \sin k_0)}$ and the normalization is given by $N=\sqrt{2 \zeta_{\pm} \sqrt {r}} 
e^{i \frac{\theta}{4}}$. The dual vectors corresponding to $|~\pm~\rangle$ are identified as 
\begin{equation}
 \overline{ \langle~\pm~|}=N^{-1}\{{\langle~L, \cal{EP}}~|+ \zeta_{\pm} \, \sqrt {r}~ \langle~L(a),{\cal{EP}}~|~e^{i \frac{\theta}{2}}\}~,
 \end{equation}
  which satisfy the orthonormalization condition,
  \begin{equation}
 \overline{\langle ~\pm~|}~\pm~\rangle=1 \quad,  \quad \overline{\langle ~\pm~|}~\mp~\rangle=0~,
 \end{equation}
 hence indicating that once we are away from the $\cal{EP}$ the standard formulation of normalization condition of  bi-orthogonal vectors given in Eq.(\ref{NC}) holds. The over line of dual vectors is chosen to emphasize the fact that these are not the same as  ${|~\pm~\rangle}^\dagger$. This expansion sets the stage for evaluating Berry phase and quantum geometric tensor around  $\cal{EP}$.\\
 {\it{{\underline{Berry phase around the $\cal{EP}$s}}:-}} We note that the eigenvalues exchange themselves ($E_{\pm}(\theta) \rightarrow E_{\mp} (\theta)$) as we go once around the $\cal{EP}$, i.e. $\theta \rightarrow \theta + 2 \pi$ and they return to themselves as we go around twice ($\theta \rightarrow \theta + 4 \pi$). The branch point geometry of energy associated with these non-Hermitian degeneracies are to be contrasted with that of Diabolic geometry \cite{diabolic}
\footnote{For a $2\times2$ hermitian Hamiltonian which is parametrized in terms of two real parameters, diabolic point is a double cone looking touching point of negative and positive energy eigenvalue surfaces obtained by plotting the two eigenvalues as a function of these two parameters.} 
Also, the eigenvectors exchange themselves up to phase $|~\pm,\theta~\rangle \rightarrow -i  |~\mp,\theta~\rangle$ as  $\theta \rightarrow \theta + 2 \pi$. One  going around twice, the states return to themselves $|~\pm,\theta~\rangle \rightarrow -~ |~\pm,\theta~\rangle$ except for the negative sign.

This negative sign represents a topological phase of $\pi$ which is considered as an hallmark of an exceptional point \cite{exotic} confirming a M$\ddot{\rm o}$bius  strip-like  topology. Base on above analysis we conclude that zero energy states can be identified with a pair of $\cal{EP}$.\\
{\it{{\underline{Quantum geometric characterization of $\cal{EP}$}}:-}} 
Hilbert space of a non-Hermitian Hamiltonian has a local gauge freedom given by  $|~\psi(x_i)~\rangle \rightarrow e^{i \alpha(x_i)}|~\psi(x_i)~\rangle ,  \overline{ \langle~\psi(x_i)~|} \rightarrow e^{-i \alpha(x_i)}~  \overline{ \langle~\psi(x_i)~|} $ where $x_i$ represents local coordinates in the space of parameters parametrizing the Hamiltonian and $\overline{ \langle~\psi(x_i)~|}$~($\neq {(|~\psi(x_i)~\rangle)^{\dagger}}$) is the dual of $|~\psi(x_i)~\rangle$  which can be expanded in an appropriate bi-orthogonal basis. Now we can define a gauge invariant derivative of $|~\psi(x_i)~\rangle$ which transforms covariantly under the gauge transformation given by  $|~D_i\psi~\rangle = |~\partial_i \psi~\rangle + |~\psi~\rangle~  \overline{ \langle~\psi~|}~\partial_i \psi~\rangle $ and a dual defined similarly denoted by $\overline{\langle~D_i\psi~|}$. This construction helps us define a gauge covariant second rank tensor which can be expressed as a sum of  symmetric and a anti-symmetric part given by  $\overline{\langle~D_i\psi~|}~D_j\psi~\rangle= (1/2) (g_{i j} + i V_{i j})$ which is analog to the corresponding formulation for the Hermitian case \cite{GPbook}.  Here the anti-symmetric part $V_{i j}$ represents the Berry curvature for the non-Hermitian case leading to the analog of the Berry phase which has both a real part (similar to Hermitian case) and an imaginary parts (corresponding to geometric dissipation which is non existent of Hermitian case) \cite{garrison}. The symmetric part defines a metric tensor which defines a notion of geometric distance between various states in the Hilbert space of the non-Hermitian Hamiltonian given by 
\begin{equation}
g_{i j}= \frac{1}{2}\{\,\overline{\langle~\partial_i\psi~|}\partial_j \psi~\rangle - \overline{ \langle~\partial_i\psi~|} \psi~\rangle  \overline{ \langle~\psi~|} \partial_j\psi~\rangle + i\leftrightarrow j \, \}~.
\label{metric}
\end{equation}
To evaluate $g_{i j}$ in the neighborhood of $\cal{EP}$ we prefer to use the Cartesian coordinate corresponding to $k-k_0=k_1 + i \, k_2$ and we reparamatrized  $|~\pm,\theta~\rangle$ as  $|~\pm,(k_1,k_2)~\rangle$. It is straightforward to show that $\partial_{k_1}{ |~\pm,(k_1,k_2)~\rangle}=-\{ 1/{4 (k-k_0)} \} |~\mp,(k_1,k_2)~\rangle$ and $\partial_{k_2}{ |~\pm,(k_1,k_2)~\rangle}=-\{ i/{4 (k-k_0)} \} |~\mp,(k_1,k_2)~\rangle$ which implies that the second term in the expression of $g_{i j}$ is identically zero while the first term leads to $g_{i j}\propto 1/{(k-k_0)^2}$. This singular behavior of the quantum metric in the neighborhood of the $\cal{EP}$ is again a hallmark of the $\cal{EP}$ \cite{brody}.\\
{\it{{\underline{Collapse of $\cal{EP}$ at the transition point}}:-}}
 Now we turn to the issue of the of phase transition. At the transition point  $\mu=1$ \cite{Kitaev}, the two zero energy solutions appear at $k_0^{+}=0$ and $k_0^{-}=i \log \{(1+\Delta)/(1-\Delta)\}$ (from Eq.(\ref{ising0})). Using  Eq.(\ref{eigenstate}) it is straight forward to check that $k_0^{-}$ indeed corresponds to an exceptional point. On the other hand at $k_0^{+}=0$,  we note that $a_{k_0}^{}=\Delta \sin k_0^{+}=0$  which leads to vanishing of the coefficient of the Jordan block itself in Eq.(\ref{CP}). 
Hence the exceptional point corresponding to $k_0^{+}$ vanishes exactly at  $\mu=1$. To understand what physics takes over the $\cal{EP}$ corresponding to $k=k_0^{+}$ at $\mu=1$, we perform an expansion of $H(k)$ around this point to leading order in $k-k_0^{+}$ which gives 
\begin{equation}
H(k) \approx -\Delta \sigma_x ( k -k_0^{+})~.
\label{transition}
\end{equation}
Therefore, the critical point is described by a $1$-$D$ Dirac type Hamiltonian representing a perfect level crossing at $k=k_0^{+}$ and hence is distinct from the $\cal{EP}$ Hamiltonian  of Jordan normal form.
It is straight forward to check that the quantum metric has no singularities at $k=k_0^{+}$ unlike an $\cal{EP}$.\\
To conclude,  in the space of complex-$k$ the quantum phase transition is marked by the conversion of one of the two $\cal{EP}$s to trivial degeneracy and then back to $\cal{EP}$ as we cross the transition. Degeneration of a pair of $\cal EP$ into one exceptional and one trivial degeneracy at the critical point of the topological phase transition unveils a new scenario quite distinct from popularly observed bifurcation of a Diabolic point into two $\cal EP$s associated with gap closing discussed in earlier studies \cite{Keck}.  This characterization of topological phase transition is one of the key results of our paper. \\
{\it{{\underline{Majorana count in terms of $\cal EP$}}:-}} From Eq.(\ref{ising0}), for  $\mu > 1$  we have  $Im (k_0^{+}) < 0, Im (k_0^{-})>0$ and for $0 \leq \mu < 1$ we have $Im (k_0^{+}) > 0, Im (k_0^{-}) > 0$ and $Im ({k_0^{-})=0}$ at $\mu=1$. Here $\Im m$ represents imaginary part.  Hence ${\text{sgn}}\{ Im (k_0^{+})\}$ (${\text{sgn}}$ is signum function) define a $Z_2$ valued  quantum number which changes from $-1$ to $+1$ across the transition. And the count of Majorana fermion pairs in our formulation is simply given by $(1/2)[{\text{sgn}}\{Im (k_0^{+})\} + {\text{sgn}} \{Im (k_0^{-})\}]$ which is zero for  $\mu > 1$ (non-topological phase) and is one for  $0 \leq \mu < 1$. \\
 {{\it {{\underline{Conclusion and discussions}}:-}}}
  In this letter we have looked for an alternative route for obtaining confirmatory signatures of topological boundary states without breaking the translational invariance by posing a boundary. We accomplished it by extending the Bloch momentum to complex planes and then looking for bound states solutions. We have successfully applied this idea to  the 1-D  p-wave superconductor and established a new topological descriptions of Majorana bound state in terms exceptional point. The natural question which comes next is; how general is this approach? Actually such an approach also works for two dimensional topologically nontrivial states of matter like the quantum Hall effect\cite{Hatsugai}) but unlike our case they did not have a exceptional point like topology. Also, recent study of graphene \cite{GEP} has used this approach and shown the existence of $\cal EP$s but they are not connected to Majorana. \\ \\
 {{\it {Acknowledgments:}}} It is a pleasure to thank Michael Berry for email correspondence, Poonam Mehta for useful discussions and Johannes Motruk for critical reading of the manuscript and comments.
\vspace{-.1in}

\bibliographystyle{apsrev}

\bibliography{references}

\begin{thebibliography}{29}
\expandafter\ifx\csname natexlab\endcsname\relax\def\natexlab#1{#1}\fi
\expandafter\ifx\csname bibnamefont\endcsname\relax
  \def\bibnamefont#1{#1}\fi
\expandafter\ifx\csname bibfnamefont\endcsname\relax
  \def\bibfnamefont#1{#1}\fi
\expandafter\ifx\csname citenamefont\endcsname\relax
  \def\citenamefont#1{#1}\fi
\expandafter\ifx\csname url\endcsname\relax
  \def\url#1{\texttt{#1}}\fi
\expandafter\ifx\csname urlprefix\endcsname\relax\def\urlprefix{URL }\fi
\providecommand{\bibinfo}[2]{#2}
\providecommand{\eprint}[2][]{\url{#2}}

\bibitem[{\citenamefont{{Wilczek}}(2012)}]{Wil}
\bibinfo{author}{\bibfnamefont{F.}~\bibnamefont{{Wilczek}}},
  \bibinfo{journal}{Nature} \textbf{\bibinfo{volume}{486}},
  \bibinfo{pages}{195} (\bibinfo{year}{2012}).

\bibitem[{\citenamefont{{Kitaev}}(2001)}]{Kitaev}
\bibinfo{author}{\bibfnamefont{A.~Y.} \bibnamefont{{Kitaev}}},
  \bibinfo{journal}{Physics Uspekhi} \textbf{\bibinfo{volume}{44}},
  \bibinfo{pages}{131} (\bibinfo{year}{2001}).

\bibitem[{\citenamefont{Lutchyn et~al.}(2010)\citenamefont{Lutchyn, Sau, and
  Das~Sarma}}]{Mftheo2}
\bibinfo{author}{\bibfnamefont{R.~M.} \bibnamefont{Lutchyn}},
  \bibinfo{author}{\bibfnamefont{J.~D.} \bibnamefont{Sau}}, \bibnamefont{and}
  \bibinfo{author}{\bibfnamefont{S.}~\bibnamefont{Das~Sarma}},
  \bibinfo{journal}{Phys. Rev. Lett.} \textbf{\bibinfo{volume}{105}},
  \bibinfo{pages}{077001} (\bibinfo{year}{2010}).

\bibitem[{\citenamefont{Oreg et~al.}(2010)\citenamefont{Oreg, Refael, and von
  Oppen}}]{Mftheo3}
\bibinfo{author}{\bibfnamefont{Y.}~\bibnamefont{Oreg}},
  \bibinfo{author}{\bibfnamefont{G.}~\bibnamefont{Refael}}, \bibnamefont{and}
  \bibinfo{author}{\bibfnamefont{F.}~\bibnamefont{von Oppen}},
  \bibinfo{journal}{Phys. Rev. Lett.} \textbf{\bibinfo{volume}{105}},
  \bibinfo{pages}{177002} (\bibinfo{year}{2010}).

\bibitem[{\citenamefont{Law et~al.}(2009)\citenamefont{Law, Lee, and
  Ng}}]{Mftheo4}
\bibinfo{author}{\bibfnamefont{K.~T.} \bibnamefont{Law}},
  \bibinfo{author}{\bibfnamefont{P.~A.} \bibnamefont{Lee}}, \bibnamefont{and}
  \bibinfo{author}{\bibfnamefont{T.~K.} \bibnamefont{Ng}},
  \bibinfo{journal}{Phys. Rev. Lett.} \textbf{\bibinfo{volume}{103}},
  \bibinfo{pages}{237001} (\bibinfo{year}{2009}).

\bibitem[{\citenamefont{{Beenakker}}(2011)}]{Mftheo5}
\bibinfo{author}{\bibfnamefont{C.~W.~J.} \bibnamefont{{Beenakker}}},
  \bibinfo{journal}{ArXiv e-prints}  (\bibinfo{year}{2011}),
  \eprint{1112.1950}.

\bibitem[{\citenamefont{{Mourik} et~al.}(2012)\citenamefont{{Mourik}, {Zuo},
  {Frolov}, {Plissard}, {Bakkers}, and {Kouwenhoven}}}]{Mfexpt0}
\bibinfo{author}{\bibfnamefont{V.}~\bibnamefont{{Mourik}}},
  \bibinfo{author}{\bibfnamefont{K.}~\bibnamefont{{Zuo}}},
  \bibinfo{author}{\bibfnamefont{S.~M.} \bibnamefont{{Frolov}}},
  \bibinfo{author}{\bibfnamefont{S.~R.} \bibnamefont{{Plissard}}},
  \bibinfo{author}{\bibfnamefont{E.~P.~A.~M.} \bibnamefont{{Bakkers}}},
  \bibnamefont{and} \bibinfo{author}{\bibfnamefont{L.~P.}
  \bibnamefont{{Kouwenhoven}}}, \bibinfo{journal}{Science}
  \textbf{\bibinfo{volume}{336}}, \bibinfo{pages}{1003} (\bibinfo{year}{2012}).

\bibitem[{\citenamefont{{Das} et~al.}(2012)\citenamefont{{Das}, {Ronen},
  {Most}, {Oreg}, {Heiblum}, and {Shtrikman}}}]{Mfexpt1}
\bibinfo{author}{\bibfnamefont{A.}~\bibnamefont{{Das}}},
  \bibinfo{author}{\bibfnamefont{Y.}~\bibnamefont{{Ronen}}},
  \bibinfo{author}{\bibfnamefont{Y.}~\bibnamefont{{Most}}},
  \bibinfo{author}{\bibfnamefont{Y.}~\bibnamefont{{Oreg}}},
  \bibinfo{author}{\bibfnamefont{M.}~\bibnamefont{{Heiblum}}},
  \bibnamefont{and}
  \bibinfo{author}{\bibfnamefont{H.}~\bibnamefont{{Shtrikman}}},
  \bibinfo{journal}{Nature Physics} \textbf{\bibinfo{volume}{8}},
  \bibinfo{pages}{887} (\bibinfo{year}{2012}).

\bibitem[{\citenamefont{{Rokhinson} et~al.}(2012)\citenamefont{{Rokhinson},
  {Liu}, and {Furdyna}}}]{Mfexpt2}
\bibinfo{author}{\bibfnamefont{L.~P.} \bibnamefont{{Rokhinson}}},
  \bibinfo{author}{\bibfnamefont{X.}~\bibnamefont{{Liu}}}, \bibnamefont{and}
  \bibinfo{author}{\bibfnamefont{J.~K.} \bibnamefont{{Furdyna}}},
  \bibinfo{journal}{Nature Physics} \textbf{\bibinfo{volume}{8}},
  \bibinfo{pages}{795} (\bibinfo{year}{2012}), \eprint{1204.4212}.

\bibitem[{\citenamefont{Heiss}(2012)}]{heiss}
\bibinfo{author}{\bibfnamefont{W.~D.} \bibnamefont{Heiss}},
  \bibinfo{journal}{Journal of Physics A: Mathematical and Theoretical}
  \textbf{\bibinfo{volume}{45}}, \bibinfo{pages}{444016}
  (\bibinfo{year}{2012}).

\bibitem[{\citenamefont{Keck et~al.}(2003)\citenamefont{Keck, Korsch, and
  Mossmann}}]{Keck}
\bibinfo{author}{\bibfnamefont{F.}~\bibnamefont{Keck}},
  \bibinfo{author}{\bibfnamefont{H.~J.} \bibnamefont{Korsch}},
  \bibnamefont{and} \bibinfo{author}{\bibfnamefont{S.}~\bibnamefont{Mossmann}},
  \bibinfo{journal}{Journal of Physics A: Mathematical and General}
  \textbf{\bibinfo{volume}{36}}, \bibinfo{pages}{2125} (\bibinfo{year}{2003}).

\bibitem[{\citenamefont{Korsch and Mossmann}(2003)}]{Korsch}
\bibinfo{author}{\bibfnamefont{H.~J.} \bibnamefont{Korsch}} \bibnamefont{and}
  \bibinfo{author}{\bibfnamefont{S.}~\bibnamefont{Mossmann}},
  \bibinfo{journal}{Journal of Physics A: Mathematical and General}
  \textbf{\bibinfo{volume}{36}}, \bibinfo{pages}{2139} (\bibinfo{year}{2003}).

\bibitem[{\citenamefont{Berry}(2004)}]{BerryNH}
\bibinfo{author}{\bibfnamefont{M.~V.} \bibnamefont{Berry}},
  \bibinfo{journal}{Czech. J. Phys.} \textbf{\bibinfo{volume}{54}},
  \bibinfo{pages}{1039} (\bibinfo{year}{2004}).

\bibitem[{Note1()}]{Note1}
Note1, \bibinfo{note}{it should noted that based on our idea of describing
  Majorana zero mode as $\protect \cal {EP}$ its application to higher
  dimensions has already been published in EPL,110, 67005 (2015) by I Mandal}.

\bibitem[{\citenamefont{Cheon et~al.}(2009)\citenamefont{Cheon, Tanaka, and
  Kim}}]{exotic}
\bibinfo{author}{\bibfnamefont{T.}~\bibnamefont{Cheon}},
  \bibinfo{author}{\bibfnamefont{A.}~\bibnamefont{Tanaka}}, \bibnamefont{and}
  \bibinfo{author}{\bibfnamefont{S.~W.} \bibnamefont{Kim}},
  \bibinfo{journal}{Physics Letters A} \textbf{\bibinfo{volume}{374}},
  \bibinfo{pages}{144} (\bibinfo{year}{2009}).

\bibitem[{\citenamefont{{Bender}}(2007)}]{Bender2}
\bibinfo{author}{\bibfnamefont{C.~M.} \bibnamefont{{Bender}}},
  \bibinfo{journal}{Reports on Progress in Physics}
  \textbf{\bibinfo{volume}{70}}, \bibinfo{pages}{947} (\bibinfo{year}{2007}).

\bibitem[{\citenamefont{Bender and Boettcher}(1998)}]{Bender1}
\bibinfo{author}{\bibfnamefont{C.~M.} \bibnamefont{Bender}} \bibnamefont{and}
  \bibinfo{author}{\bibfnamefont{S.}~\bibnamefont{Boettcher}},
  \bibinfo{journal}{Phys. Rev. Lett.} \textbf{\bibinfo{volume}{80}},
  \bibinfo{pages}{5243} (\bibinfo{year}{1998}).

\bibitem[{\citenamefont{Hatsugai}(1997)}]{Hatsugai}
\bibinfo{author}{\bibfnamefont{Y.}~\bibnamefont{Hatsugai}},
  \bibinfo{journal}{Journal of Physics: Condensed Matter}
  \textbf{\bibinfo{volume}{9}}, \bibinfo{pages}{2507} (\bibinfo{year}{1997}).

\bibitem[{\citenamefont{Horn and Jonson}(1985)}]{jordan}
\bibinfo{author}{\bibfnamefont{R.~A.} \bibnamefont{Horn}} \bibnamefont{and}
  \bibinfo{author}{\bibfnamefont{C.~R.} \bibnamefont{Jonson}},
  \emph{\bibinfo{title}{Matrix Analysis}} (\bibinfo{publisher}{Cambridge
  University Press}, \bibinfo{year}{1985}).

\bibitem[{Note2()}]{Note2}
Note2, \bibinfo{note}{for the unphysical case of negative $\mu $, the
  topological window for $\mu $ is given by $-1< \mu \leq 0$ and in this window
  the correct identification of Majorana solution corresponds to the $A_k=0$
  case.}

\bibitem[{\citenamefont{{G{\"u}nther} et~al.}(2007)\citenamefont{{G{\"u}nther},
  {Rotter}, and {Samsonov}}}]{rotter}
\bibinfo{author}{\bibfnamefont{U.}~\bibnamefont{{G{\"u}nther}}},
  \bibinfo{author}{\bibfnamefont{I.}~\bibnamefont{{Rotter}}}, \bibnamefont{and}
  \bibinfo{author}{\bibfnamefont{B.~F.} \bibnamefont{{Samsonov}}},
  \bibinfo{journal}{Journal of Physics A Mathematical General}
  \textbf{\bibinfo{volume}{40}}, \bibinfo{pages}{8815} (\bibinfo{year}{2007}),
  \eprint{0704.1291}.

\bibitem[{\citenamefont{Berry and Uzdin}(2011)}]{BerryNH1}
\bibinfo{author}{\bibfnamefont{M.~V.} \bibnamefont{Berry}} \bibnamefont{and}
  \bibinfo{author}{\bibfnamefont{R.}~\bibnamefont{Uzdin}},
  \bibinfo{journal}{Journal of Physics A: Mathematical and Theoretical}
  \textbf{\bibinfo{volume}{44}}, \bibinfo{pages}{435303}
  (\bibinfo{year}{2011}).

\bibitem[{\citenamefont{Berry}(2011)}]{BerryNH2}
\bibinfo{author}{\bibfnamefont{M.~V.} \bibnamefont{Berry}},
  \bibinfo{journal}{Journal of Optics} \textbf{\bibinfo{volume}{13}},
  \bibinfo{pages}{115701} (\bibinfo{year}{2011}).

\bibitem[{\citenamefont{Garrison and Wright}(1988)}]{garrison}
\bibinfo{author}{\bibfnamefont{J.~C.} \bibnamefont{Garrison}} \bibnamefont{and}
  \bibinfo{author}{\bibfnamefont{E.~M.} \bibnamefont{Wright}},
  \bibinfo{journal}{Physics Letters A} \textbf{\bibinfo{volume}{128}},
  \bibinfo{pages}{177 } (\bibinfo{year}{1988}).

\bibitem[{\citenamefont{Brody and Graefe}(2013)}]{brody}
\bibinfo{author}{\bibfnamefont{D.~C.} \bibnamefont{Brody}} \bibnamefont{and}
  \bibinfo{author}{\bibfnamefont{E.-M.} \bibnamefont{Graefe}},
  \bibinfo{journal}{Entropy} \textbf{\bibinfo{volume}{15}},
  \bibinfo{pages}{3361} (\bibinfo{year}{2013}).

\bibitem[{\citenamefont{Berry and Wilkinson}(1984)}]{diabolic}
\bibinfo{author}{\bibfnamefont{M.~V.} \bibnamefont{Berry}} \bibnamefont{and}
  \bibinfo{author}{\bibfnamefont{M.}~\bibnamefont{Wilkinson}},
  \bibinfo{journal}{Proc. R. Soc. Lond. A} \textbf{\bibinfo{volume}{392}},
  \bibinfo{pages}{15} (\bibinfo{year}{1984}).

\bibitem[{Note3()}]{Note3}
Note3, \bibinfo{note}{for a $2\times 2$ hermitian Hamiltonian which is
  parametrized in terms of two real parameters, diabolic point is a double cone
  looking touching point of negative and positive energy eigenvalue surfaces
  obtained by plotting the two eigenvalues as a function of these two
  parameters.}

\bibitem[{\citenamefont{Shapere and Wilczek}(1989)}]{GPbook}
\bibinfo{author}{\bibfnamefont{A.}~\bibnamefont{Shapere}} \bibnamefont{and}
  \bibinfo{author}{\bibfnamefont{F.}~\bibnamefont{Wilczek}},
  \emph{\bibinfo{title}{Geometric Phases in Physics}}
  (\bibinfo{publisher}{World Scientific, Singapore}, \bibinfo{year}{1989}).

\bibitem[{\citenamefont{Fagotti et~al.}(2011)\citenamefont{Fagotti, Bonati,
  Logoteta, Marconcini, and Macucci}}]{GEP}
\bibinfo{author}{\bibfnamefont{M.}~\bibnamefont{Fagotti}},
  \bibinfo{author}{\bibfnamefont{C.}~\bibnamefont{Bonati}},
  \bibinfo{author}{\bibfnamefont{D.}~\bibnamefont{Logoteta}},
  \bibinfo{author}{\bibfnamefont{P.}~\bibnamefont{Marconcini}},
  \bibnamefont{and} \bibinfo{author}{\bibfnamefont{M.}~\bibnamefont{Macucci}},
  \bibinfo{journal}{Phys. Rev. B} \textbf{\bibinfo{volume}{83}},
  \bibinfo{pages}{241406} (\bibinfo{year}{2011}).

\end{thebibliography}

\end{document}